%Paper: hep-ph/9305204
%From: BEST@sns.ias.edu
%Date: 03 May 1993 14:46:29 -0400 (EDT)
%Date (revised): 02 Jun 1993 09:55:27 -0400 (EDT)

\def\doublespace{\baselineskip=2\normalbaselineskip}
\def\ltorder{\mathrel{\raise.3ex\hbox{$<$}\mkern-14mu
             \lower0.6ex\hbox{$\sim$}}}
\def\gtorder{\mathrel{\raise.3ex\hbox{$>$}\mkern-14mu
             \lower0.6ex\hbox{$\sim$}}}

\def\Rf{\baselineskip=2\normalbaselineskip
\parindent=0pt \medskip\hangindent=3pc
\hangafter=1 }
\def\tcar{\futurelet\next\testnextcar}
\def\testnextcar{\ifhmode\ifcat\next.\else\ \fi\fi}
\def\bnue{$\bar\nu_e$\tcar}
\def\bnum{$\bar\nu_\mu$\tcar}
\doublespace
\centerline{\bf  IS LARGE LEPTON MIXING EXCLUDED?}
\vskip 1cm
\centerline {\bf Alexei Yu. Smirnov$^{(1),(3),(4)}$,
                 David N. Spergel$^{(2)}$ and John N. Bahcall$^{(1)}$}
\centerline{\it $^{(1)}$ Institute for Advanced Study, Princeton, NJ 08540,
USA}
\centerline{\it $^{(2)}$ Princeton University Observatory, Princeton,
NJ 08540, USA}
\centerline{\it $^{(3)}$ International Centre for Theoretical Physics,
34100 Trieste, Italy}
\centerline{\it $^{(4)}$ Institute for Nuclear Research, 117312 Moscow, Russia}
\vskip 1cm
\centerline{\bf Abstract}

\doublespace

The original \bnum -
(or $\bar{\nu}_{\tau}$-)
energy spectrum from the gravitational collapse of a star
has a larger average energy than the spectrum for \bnue
since the opacity of \bnue exeeds that of \bnum (or $\nu_{\tau}$).
Flavor neutrino conversion, \bnue $\leftrightarrow$ \bnum,
induced by lepton mixing results
in partial permutation of the original \bnue and \bnum spectra.
An upper bound on the permutation factor,
 $p \leq 0.35$ (99$\%$ CL)
is derived using  the data from SN1987A and
the different models of the neutrino burst.
The
relation between  the permutation factor and the vacuum
mixing angle is established, which leads to the
upper bound on this angle.  The excluded region,
$\sin^2 2\theta > 0.7 - 0.9$, covers the regions  of
large mixing angle solutions of the solar neutrino problem:
``just-so" and, partly, MSW, as well as
part of region of $\nu_{e} - \nu_{\mu}$
oscillation space which could be responsible for the
atmospheric muon neutrino deficit.  These limits are sensitive to the
predicted neutrino spectrum and can be strengthened as supernova
models improve.

\vskip 1cm
\vfill\eject
\noindent
{ \bf I. Introduction}

There are several hints that lepton mixing does exist and
might even be much bigger than that in the quark sector.
Solar neutrino data [1] can be reconciled with predictions of the
Standard Solar Model [2] by long length vacuum oscillations
(``just-so" solution) [3]. The required values of neutrino
mixing angle, $\theta$, and masses squared difference, $\Delta m^2$,
 are: $\sin^2 2\theta = 0.85 - 1.0$,
$\Delta m^2 = (0.8 - 1.1) \cdot 10^{-10}~ {\rm eV}^2$ [4].
The solar neutrino problem can be solved also by resonant flavor
conversion, the MSW-effect [5]. For the MSW solution,
the data single out two regions of neutrino
parameters, one of which involves large  mixing angles:
$\sin^2 2\theta = 0.6 - 0.9$, at
$\Delta m^2 = (10^{-7} - 10^{-5})~ {\rm eV}^2$ [6].
The deficit of the muon neutrinos in the atmospheric neutrino flux
can be explained by $\nu_{\mu} - \nu_e$ oscillations with parameters [7]:
$\sin^2 2\theta = 0.5 - 0.9$,
$\Delta m^2 = (10^{-3} - 10^{-2})~ {\rm eV}^2$ (see fig.3).

On the other hand it has been argued that mixing in the lepton
sector can be ``naturally" large. In particular, large lepton
mixing may appear in models with radiative generation of
the neutrino masses (Zee-mechanism [8], see [9] for review).
In the ``see-saw" mechanism some configurations of
mass matrices result in large mixing angles (see, e.g., [10]);
the ``see-saw" enhancement of lepton mixing may take
place at definite conditions (strong mass hierarchy in Majorana
mass sector, or definite symmetry of the
 majorana mass matrix and mass degeneration of the right-handed
neutrino components [11]).

Large  lepton mixing can be generated by some interactions at the
Planck scale, which result in nonrenormalizable terms
of the type $(\alpha_{ij}/M_{Pl})\cdot l_{i}^T l_{j} H^+ H$
[12,13]. Here $l_{i}$ (i = $e, \mu, \tau$) are the lepton doublets of
definite flavor, H is the Higgs doublet, and $M_{Pl}$ is the Planck mass.
At $\alpha \approx 1$,
these terms generate the neutrino masses $m_{ij} = \langle H \rangle ^2
/M_{Pl} \approx 10^{-5}$ eV, which gives $\Delta m^{2}$
in the region of ``just-so" solutions. Furthermore, it was argued in [13]
that the ``Planck-scale interaction" related to gravity
does not respect lepton number,
and moreover  all coupling constants in the flavor
basis have the same value $\alpha_{ij} = \alpha_{0}$[13].
The corresponding mass matrix has
all elements equal to each other.
In this case,
the electron neutrino mixes with only one state, namely, with the combination
$(\nu_{\mu} - \nu_{\tau})/ \sqrt {2}$, and the mixing parameter
is $\sin^2 2\theta = 8/9$, i.e.
precisely in the ``just-so" region.
Although there is no real model for the ``Planck-scale interaction"
the coincidence of
parameters is remarkable.

In this paper we will discuss  the limits on large lepton
mixing that could be obtained using the observational data from
the supernova SN1987A [14, 15, 16].

The effects of lepton mixing on the neutrino fluxes from
gravitational collapses of stars have been widely discussed
[17 - 22, 5]. In particular, it was noted that large flavor
mixing results in a
significant distortion
of the \bnue - spectra at the Earth; the
appearance of a high-energy tail, and thus the increase of the
average energy of the detected
events relative to the no mixing case
are expected [17, 19, 20]. Comparing the spectra with and
without mixing effects it was remarked in [19] that Kamiokande data
seem to disfavour $\theta > 50^0$.  At
large mixing angles, the oscillations in the matter of the
Earth result in different signals
in Kamiokande and IMB detectors;
this could explain the more energetic
spectrum seen by IMB [20].
Here we refine the consideration  of the large mixing  effects
to obtain statistically significant upper bounds on the mixing angle by
make of use
of the existing data from SN1987A.

\vskip 0.8cm
\noindent
{ \bf II. Permutation of \bnue and \bnum spectra.
Permutation factor. }

Consider the influence of transitions $\bar{\nu}_e \leftrightarrow
\bar{\nu}_{\mu}$ on the \bnue - energy spectrum. Since
\bnum and $\bar{\nu}_{\tau}$
 have, to high accuracy,
the same production and detection properties, the results
will be the same for transitions to $\bar{\nu}_{\tau}$ or to any
combination of \bnum and $\bar{\nu}_{\tau}$.
(This remark applies also for transitions into
$\nu_{\mu}$ and $\nu_{\tau}$).
We will comment on three-neutrino mixing latter (in Sect.~IV),
although many cases can be
reduced to two neutrino mixing.

Let $F_0(\bar\nu_e)$ and $F_0(\bar\nu_\mu)$ be the original
\bnue -, and \bnum - spectra,  and let $p$ be the probability
of a \bnue $\rightarrow$ \bnum transition on the way from a core
of collapsing star to the detector. Since the
\bnue -, and \bnum - spectra emitted
by  neutrinospheres are incoherent, the \bnue flux in
the detector can be written as
$$
F(\bar\nu_e)~=~(1 - p) \cdot F_0(\bar\nu_e)~+~p \cdot
F_0(\bar\nu_\mu).
\eqno(1)
$$
Obviously, there is no observable
effect when the original spectra
are the same: $F_0(\bar\nu_e)~=~$ $F_0(\bar\nu_\mu)$.

The energy spectra of \bnue 's and \bnum 's that are emitted from
the core of a collapsing star are different: the
\bnue - spectrum has a mean energy that is 1.5 - 2 times smaller than
that of the \bnum -
spectrum. This general feature
follows from the fact that \bnue interacts with matter
more strongly  than \bnum does; neutral current scattering
and charged current absorption on protons,
$\bar{\nu}_{e} + p \rightarrow n + e^+$,  are allowed
for \bnue but not for \bnum. Also, due to the charge current
interaction, the cross-section of \bnue-scattering on
electrons is larger than that for \bnum.
Therefore \bnue's encounter a larger
opacity and consequently are emitted from more
external and colder layers of the star. This essentially
model-independent feature plays a key role in our
determination of the maximum allowed mixing angles. Another
crucial point is that the cross section of the detection
reaction, $\bar{\nu}_e + p \rightarrow e^+ + n$, is approximately
proportional to the neutrino energy squared. Therefore
even a small permutation (or admixture of a higher energy
spectrum)
can result in an appreciable effect.

The  transformation factor, $p$, depends on the energy of the
neutrino, the mass splitting, the vacuum mixing angle, and
the density profiles of the supernova and the Earth.
However, for most of the interesting mass range, $p$
is independent of energy.  For $\Delta m^2 \sim 10^{-6}$ eV$^2$,
the energy dependence is important and we average the
permutation factor over the energy distribution.
If $p
= 1$ (complete transformation), the detected \bnue - spectrum
will coincide with the original \bnum - spectrum:
$F(\bar\nu_e) = F_0(\bar\nu_\mu)$ and,  vice versa, the
final \bnum  will coincide with original spectra of \bnue.
The
spectra permute and we will call the average probability $p$
the permutation factor.
If $p < 1$, only partial permutation takes place and the
final \bnue-energy spectrum will be
a mixture of the two original \bnue and \bnum spectra.

Fig.1 depicts the expected cumulative energy spectrum of the events
in Kamiokande-II and IMB detectors for
different values of $p$.  The parameters of a ``conventional
neutrino burst" [23, 24, 25] have been used.
As many authors have concluded previously, the observed
energy spectrum from SN1987A
is in reasonable agreement with that calculated
without any neutrino transformations. Fig.1 shows that
the \bnue $\rightarrow$ \bnum transition produces unobserved
 high energy events.
We use this result to exclude large
values of $p$.

\vskip 0.8cm

\noindent
{ \bf III. Upper bounds on the permutation factor
from SN1987A data}

We will compare the shapes of the predicted time-integrated
energy spectra for
different values of $p$ with the observed energy distribution.
The original \bnue-, and \bnum- spectra are approximated by the
modified Fermi-Dirac spectrum [26 - 32]:
$$
{d E^{tot} \over d E } = {A E^3 \over e^{({E \over T} - \eta)} + 1}~,
$$
where $A, T, \eta$ are the fit parameters.
The modification is related to the fact that the emitted spectra
are superpositions of thermal fluxes (in general, Fermi-Dirac spectra with
nonzero chemical potentials) from different thermalization
spheres. These spectra are further modified by scattering
and absorption above the thermalization spheres
and by the integration over the
neutrino burst time (the effective temperatures are changed during the
burst).
Instead of $T$,
we fix the average energy of the spectrum, $\bar{E}$, :
$$
\bar{E} = {E^{tot} \over \int {d E^{tot} \over d E} {d E \over E}},
$$
where $E^{tot} \equiv \int {d E^{tot} \over d E} d E$
is the total emitted energy in a given neutrino type.
The distortion parameter, $\eta$, has the effect of a ``chemical
potential".
The parameters of the  time-integrated spectra
that we use in our analysis are:
the average energies of the electron antineutrino, $\bar{E}_e$,
and the muon antineutrino, $\bar{E}_{\mu}$,
as well as the ratio
of the total energies emitted in \bnue 's and \bnum 's:
$r  \equiv {E^{tot}(\bar{\nu}_{\mu}) \over E^{tot}(\bar{\nu}_e)}$.
The absolute value of the
total energy carried away by neutrinos is  eliminated by normalization;
the total number of calculated  events
is  constant and equal to the observed value, $N = 20$.

The cumulative energy spectra of
observed and calculated neutrino events are compared by the
Kholmogorov-Smirnov test, which allows us to set non-parametric upper
limits on $p$ at a definite confidence level for different values
of the original spectra parameters (see fig.2).
As is apparent from fig.2,  the inferred upper bound  depends
most strongly on
the average muon neutrino energy, $\bar{E}_{\mu}$;
the main difference between the calculated and the observed
 spectra comes from the high
energy region for which the calculated events
are caused by \bnum  converted to \bnue.
If $\bar{E}_{\mu} < 6.5$ MeV, then the bound is $p < 0.5$
and no strong limit can be obtained for the antineutrino
channel (see Sect.~IV). The bounds depend weakly
on the total fluence emitted in  \bnum  \ (fig. 2a). For example, at
$\bar{E}_{\mu} = 22$~MeV, one has $p$ = 0.30, 0.34, and 0.39 for
$r$ = 1.2, 1.0, and 0.8 respectively. The bounds depend
rather weakly ($5 \%$ change) on $\bar{E}_{e}$ in the most reliable region
of values 12 - 15 MeV (fig. 2b). At larger or smaller energies,
the limits become artificially strong due to the general disagreement
of the predictions  and the data even without the permutation effect.
The bounds are sensitive to the shape of the original spectra (fig. 2c).
The more pinched the spectra (bigger $\eta$), the stronger the suppression of
the number of high energy events, and, consequently, the weaker the
restrictions. For fixed $\bar{E}_{\mu}$, the dependence of bounds on
$\eta$ is stronger for smaller energies $\bar{E}_{\mu}$.
At $\bar{E}_{\mu} = 22$ MeV, the increase of $\eta$ from $0$
(pure Fermi-Dirac spectra) to $3$ results in the increase of $p$
by $15 \%$. There is a strong dependence of the inferred limits on
the assumed value of the distortion parameter of the electron antineutrino
spectrum, $\eta_e$. A decrease of $\eta_e$ results in an increase of the
number of high-energy events induced by \bnue 's and therefore strengthens
the limit on $p$.
The limits on $p$ at different confidence levels
are shown in fig. 2e. At 95$\%$ CL, a significant limit exists
even for  $\bar{E}_{\mu} = 17 - 18$ MeV. At 99.9$\%$ CL, a significant
limit can be established only for $\bar{E}_{\mu} > 22 - 23$ MeV.
At the representative value of energy
$\bar{E}_{\mu} > 22 - 24$ MeV, the $2\sigma$-limit is 35 $\%$
stronger than $3\sigma$ limit.

Baksan data [16] could be included in the analysis.
In the  Kamiokande and IMB time interval, five events
were detected at Baksan and some of these events could be related to
the neutrino burst from SN1987A. To derive an
upper bound including the Baksan data, one could use the two
most energetic Baksan events:
23 and 20 MeV
(corresponding approximately to the expected
number of events  given the number of events observed in
the bigger Kamiokande and IMB
detectors). This would
strengthen the upper bound by of order ten percent.

Since we use data from SN1987A, the model of collapse and
therefore the integral characteristics of the
neutrino burst
can in principle be restricted further by using
information on the progenitor and the observed
properties of light curve  of SN1987A.
The available
data suggest a  mass
of the iron core  [26] $M_{Fe} = (1.3 - 1.6) M_{\odot}$,
and, consequently, a total energy carried away
by neutrinos of $E_{tot} = (2 - 4)\cdot 10^{53}$ ergs.
The time interval of neutrino emission, $\Delta t \approx 13$ s, is in a good
agreement with the expected value, further  indicating
the basic  correctness of
the conventional picture of  neutrino transport.
The observed neutrino energies versus time suggest that the
average energy decreases with time, consistent with idea
that neutrinos
are emitted in the cooling stage.

In Table I, the principal
parameters of
different models [24 - 32] of neutrino bursts
which satisfy the above conditions
are presented and
the upper bounds
on $p$ are given in accordance with fig.2.
The restrictions:
$$
p \leq \left\{\matrix{0.35 \hfill&99\% \hfill~,\cr
                      0.23 \hfill&95\% \hfill~,\cr}\right.
\eqno(2)
$$
can be considered as upper bounds in a representative supernova neutrino
burst model.

One comment is in order. The difference between the
\bnue -,  and \bnum - spectra is determined by the difference
in interactions as well as by the structure of the star, i.e. the
density, temperature, and lepton-number profiles. The latter
in turn depends on the nuclear equation of state (EOS). A soft EOS
results in the creation of a hot and compact protoneutron star,
whereas a stiff EOS produces a colder and more expanded
central object with smaller temperatures and a smaller gradient of
temperature [33]. As a result, one  expects smaller energies
of \bnum  in the model with a stiff EOS.  In [33],
a very stiff EOS by Wolff [34] was used and the average
energies $\bar{E}_e  \approx 12$ MeV and $\bar{E}_{\mu} = 14$ MeV
were obtained. This small difference in average energies probably indicates
only the direction of a trend rather than a self consistent
numerical result. Indeed, the
model by Mayle and Wilson [24] at t = 0.4 s
after the bounce was used as the initial condition.
This model is based on a softer EOS, so that the the
calculation described in [33]  requires a non-physical
change of the EOS at 0.4 s.
The parameters at 0.4 s were adjusted to obtain
the hydrostatic configuration, whereas in the original Mayle-Wilson
   model at  t = 0.4 s the
star is still in the dynamical phase.
In [26], no strong difference of the properties of the neutrino
burst were obtained between a soft and a stiff EOS.
There is an additional reason for regarding the results of [33] with
caution.  The ``flux-limited
diffusion method" was used to describe the neutrino transport,
and the \bnue -, \bnum - energy distributions obtained
are appreciably wider than Fermi-Dirac spectra. In particular,
the calculated \bnum - spectrum can be
approximated by a  Maxwell-Boltzmann distribution ($\eta \rightarrow
-\infty$). These features [33] are in
contradiction with other results obtained by the same
method [26 - 30], as well as with results of a physically more correct
method based on Monte-Carlo simulations [31, 32].
It is of great importance to calculate a self-consistent supernova
model with the same stiff EOS [34] at all stages and to check
whether such a model fits the SN1987A data
(including the neutrino luminosities and the duration of the
neutrino burst).

\vskip 0.8cm

\noindent
{ \bf IV. Permutation factors and lepton mixing}.

We consider in this section the  propagation of neutrinos
from the core of
a star to  detectors on Earth and determine
the relations between the permutation factor, $p$, and the
vacuum mixing angle, $\theta$.
We assume for most of this section that
the admixture to $\nu_{e}$ and  \bnue of the light mass
component is larger than that of the heavy component.  In this case,
the \bnue $\leftrightarrow$ \bnum channel is nonresonant.
(Matter resonance  takes place in the neutrino
channel, $\nu_e\leftrightarrow\nu_\mu$, as it is
implied by the MSW solution to
the $\nu_\odot$-problem. We will comment on the opposite case
at the end of this section.)

For the nonresonant channel, \bnue--\bnum, the mixing angle
in matter, $\theta_m$, is always
smaller than the  angle in vacuum:
$$
\sin^2 2\theta_m~=~{\tan^2 2\theta\over ({\rho\over\rho_R} + 1)^2 +
\tan^2 2\theta}~,
\eqno(3)
$$
Here $\rho$ is the density, $m_N$ is the nucleon mass, and
$$
\rho_R~=~{m_N \Delta m^2 \cos 2\theta\over 2\sqrt{2} G_F~Y_e~E}
\eqno(4)
$$
is the resonant density for the {\it neutrino} channel
($G_F$ is the Fermi constant, $Y_e$ is the number of electrons
per nucleon).

For values of neutrino parameters of interest, i.e., $\Delta m^2
\ltorder~10^{-2}~{\rm eV^2}~,~E \gtorder~10~{\rm MeV}$, the resonant
density, $\rho_R~\ltorder~10^4~{\rm
g/cm}^3$,
is much smaller than the density at the neutrino production point,
$\rho_0~\simeq~10^{12}~{\rm g/cm}^3$.  Therefore
the initial mixing is strongly suppressed:
$\sin^2 2\theta^0_m~\approx~\tan^2 2\theta \cdot
\left({\rho_R\over\rho_0}\right)^2$
and the initial neutrino
state practically coincides with eigenstate of the instantaneous
Hamiltonian of the neutrino system, $\bar\nu_{1m}$,:
$\bar\nu (t = 0) \equiv \bar\nu_e \cong \bar\nu_{1m}$.
Further evolution of
this state is determined by the
adiabaticity condition [5].
(If this condition
 is fulfilled, the transitions of the eigenstates,
$\bar\nu_{1m} \leftrightarrow \bar\nu_{2m}$,
can be neglected).
The adiabaticity
condition reads:  $\kappa \ll 1$, where
$\kappa~\equiv~{d \theta_m/dr\over \Delta H}$
is the adiabaticity parameter [5].
Here $\Delta H$ is the energy splitting
between eigenvalues of the Hamiltonian,
$\Delta H \equiv E(\bar\nu_{1m}) - E(\bar\nu_{2m}$). The adiabaticity parameter
can be written in the following form:
$$
\kappa~=~{\sin^3 2\theta_m \over \sin^2 2\theta} \cdot {l^2_\nu\over
4\pi~l_\rho \cdot l_0}~,
\eqno(5)
$$
where $l_\nu~=~4\pi E/\Delta m^2$ is the oscillation length in vacuum,
$l_0~=~2\pi m_N /\sqrt{2} G_F \rho Y_e$ is the refraction length, and
$l_\rho~\equiv~\rho/(d\rho/dr)$ is the typical density
scale height.

Since $\sin^3 2\theta_m \propto 1/\rho^3$ at $\rho \gg \rho_R$ and
$l_{0}^{-1} \propto \rho$, the parameter $\kappa$
is small at large densities, and the adiabaticity condition
is fulfilled in the early stage of neutrino propagation.
When the density
decreases, $\kappa$ first increases as $1/\rho^2$,
reaches a maximum value at
$\rho_m  \sim \rho_R$
($\rho_m /\rho_R = {3 \over 4} y - {1 \over 4}$,
where $y \equiv  \sqrt {1 + {8 \over 9} \tan^2 2\theta}$),
 and then decreases again as $\rho$.
For $\rho = \rho_m$ we get from (5):
$$
\kappa_{m}~=~{f(\theta) \over 4\pi}  \cdot {l_\nu \over l_\rho},
\eqno(6)
$$
where
$$
f(\theta) = {16 \over 9}\cdot{\tan ^2 2\theta \over  \sin 2 \theta}
\cdot {y - {1 \over 3} \over \left[(y + 1)^2 +
{16 \over 9} \tan^2 2\theta \right]^{3 \over 2}}.
$$
The function $f(\theta)$ increases from $\approx 0.09$
at $\sin^2 2\theta~=~0.3$, to $\approx 0.28$
at $\sin^2 2\theta = 0.95$; ($f = 2$ at
$\sin^2 2\theta~=~1)$.  Substituting a typical value
$f(\theta) = 0.2$ into (6),  we find
$$
\kappa < \kappa_m~\leq~6 \cdot 10^{-10}\left({1~{\rm eV}^2 \over
\Delta m^2}\right)\left({E \over 10~{\rm
MeV}}\right)\left({R_\odot\over l_\rho}\right)~.
\eqno(7)
$$
For $l_\rho = R_\odot$ one obtains from (7)
that $\kappa_R = 1$ (strong adiabaticity violation at
resonance densities) at
$$
\Delta m^2_a~=~\left\{\matrix{6 \cdot 10^{-10}~{\rm eV}^2\hfill&(E =
10~{\rm MeV})\hfill\cr
3 \cdot 10^{-9}~{\rm eV}^2\hfill&(E = 50~{\rm MeV})\hfill\cr}\right.~.
\eqno(8)
$$
Note that
$\l_{\rho}$ may change from $0.1~ R_{\odot}$ in the region of
large densities, $\rho \sim 10^4$ g/cm$^3$, to $(3 - 4)\cdot R_{\odot}$  at
small densities, $\rho \sim 10^{-4}$ g/cm$^3$ [35].

The mass $\Delta m^2_a$ defines two extreme cases.
{\it Adiabatic} case: $\Delta m^2 \gg \Delta m^2_a$;
the adiabaticity condition is fulfilled everywhere in the
star. {\it Nonadiabatic} case: $\Delta m^2 \ll \Delta m^2_a$.
Here the adiabaticity is strongly broken in the region
around $\rho_R$, where the mixing angle varies from
$\theta_m \approx 0$ to $\theta_m \approx \theta$.
As we will see, the dynamics of propagation in these
extreme cases is simple and the results are essentially
independent of the  density distribution in the star.
Moreover the permutation factor is practically
independent of neutrino energy. Fortunately, the
$\Delta m^2$ regions of interest fit these two extreme
cases. The atmospheric neutrino region as well as
large mixing MSW-solutions are in the adiabatic
domain; the ``just-so" solution lies in the nonadiabatic domain.

1). In the {\it adiabatic case}
the neutrino state which is produced as
$\bar\nu_e~\cong~\bar\nu_{1m}$,  will everywhere practically
coincide with
$\bar\nu_{1m}$ since there are no
$\bar\nu_{1m} \leftrightarrow \bar\nu_{2m}$ transitions.
So the neutrino leaves the star
as $\bar\nu_{1m} (\rho = 0)$,
which is the state with definite mass $\bar\nu_1$.
No oscillations will take place on the way from the
star to the Earth and the neutrino state arriving at the Earth will be
$\bar\nu_1$.  Consequently, the probability of \bnue$\to$\bnum
transition (permutation factor) in this case equals
$p_a~=~\vert \langle \bar\nu_\mu\vert\bar\nu_1
\rangle\vert^2~=~\sin^2 \theta$ (see also [19]).

In the region of mass squared difference $\Delta m^2 =
(10^{-4} - 10^{-7})~ {\rm eV}^2$, the permutation factor
must be corrected for the effect of neutrino oscillations
inside the Earth.
 Neutrino trajectories from SN1987A to terrestrial detectors
lie in the mantle of the Earth, where the density
changes rather slowly.
Therefore, to a good approximation, one can consider  the
Earth-matter effect as neutrino oscillations in matter with
constant density ($\rho_{IMB} = 4.6$ g/cm$^{3}$ for IMB and
$\rho_{K} = 3.4$ g/cm$^3$ for Kamiokande-II).
Neutrinos
arrive at the Earth as two incoherent beams: $\nu_{1}$ - flux
with energy spectrum $F_0(\bar\nu_e)$ and $\nu_{2}$  with energy
spectrum $F_0(\bar\nu_\mu)$. Considering then the $\nu_{1} - \nu_{2}$
oscillations in the matter of the Earth, one finds the permutation factor
$$
p_a = \sin^2 \theta - \sin 2\theta_m \cdot \sin 2(\theta - \theta_m)
\cdot \sin^2 {\pi x \over l_m}~,
\eqno(9)
$$
where $\theta_{m} = \theta_m (\rho_i, E/\Delta m^2, \theta)$ (i = IMB or K)
 is the mixing angle in the matter of the Earth,
$x$ is the length of the neutrino
trajectory inside the Earth ($x_{K} = 3.9 \cdot 10^8$ cm,
$x_{IMB} = 8.4 \cdot 10^8$ cm for
Kamiokande and IMB detectors respectively),
and
$$
l_m = l_0 \cdot \left[ \left( 1 + {\rho_R \over \rho}\right)^2
+ \left({\rho_R \over \rho}\right)^2
\tan^2 2\theta\right]^{-{1 \over 2}}
\eqno(10)
$$
is the oscillation length in matter.
Here $\rho_R = \rho_R(\rho_i, E/\Delta m^2, \theta )$ is defined in
Eq.(4), $Y_e \approx 0.5$. The first term in the right hand side of
Eq.(9)
corresponds
to the adiabatic result without the Earth effect.
The second term
is the Earth correction.  Since for the nonresonant channel
$\theta >\theta_m$, this
second term  is always negative. Therefore, the Earth matter effect weakens
the permutation and relaxes  the restriction  on mixing.
The oscillation length  is always smaller than the refraction
length. Moreover, at small ${E \over \Delta m^{2}}$ (big $\Delta m^{2}$),
the oscillation length is much smaller than $l_0$.

The second term  in (9) is an oscillating
function of $x$ as well as $E/\Delta m^2$. The amplitude of
the oscillations,
$\sin 2\theta_m \cdot \sin 2(\theta - \theta_m)$,
reaches the maximal value, $\sin^2 \theta$, at
$\theta_{m} = \theta /2$. If at this point one has
$x/l_m = \pi \cdot n$ (n is integer), then the Earth effect completely
compensates the effect in the star and
$p_a = 0$. The condition for the amplitude of the correction
to be a maximum,
which can be written as
$\rho_R (E/ \Delta m^2) = \rho \cdot \cos 2 \theta$
, defines the $\Delta m^2$-region of strong Earth matter effect.
Taking into account that the interval of neutrino
energies of interest is 10 - 50 MeV, we
obtain that this region extends
over three orders of magnitude around $\Delta m^2 \approx 10^{-5}$ eV$^2$:
$\Delta m^2 = (10^{-7} - 10^{-4})~ {\rm eV}^2$.
At $\rho_R \gg \rho \cdot \cos 2 \theta$
and $\rho_{R} \ll \rho \cdot \cos 2 \theta$, the matter mixing angle
is respectively $\approx \theta$ or $\approx 0$ and therefore
the correction is negligibly small.

In the region $\Delta m^2 > 10^{-5}$ eV$^2$, the
correction is a rapidly oscillating function
of the neutrino energy.  One can average over these oscillations,
by integrating over the neutrino distribution function to yield
an average $\bar p$, which is used in Figure 3. Here $\theta_m =
\theta_m (\Delta m^2, \bar{E}, \bar{\rho})$, where
$\bar{E} = (\bar{E}_e + \bar{E}_{\mu})/2 \approx 20$ MeV
and $\bar{\rho} = (\rho_K + \rho_{IMB})/2 \approx 4.0~ {\rm g/cm}^3$.
At $\Delta m^2 \leq 10^{-5}~ {\rm eV}^2$,
the approximation $p \approx constant$
is not valid  and one must compare directly the observed
 distribution and the predicted one with an energy-dependent
oscillation factor. In this case the Earth
effect strongly depends on neutrino energy and is different
for different detectors. One can use this feature
to explain some difference in the energy distributions
of the Kamiokande and the IMB signals [20].
Fig. 4 depicts the upper bounds on  $\sin^2 2\theta$
obtained with neutrino spectra from [25].

2). {\it Nonadiabatic case}. If $\kappa \gg 1$,
neutrinos propagate nonadiabatically
in the region of strong
change of the mixing angle  ($\rho \sim \rho_R$).
The adiabaticity starts to be broken at $\rho \gg
\rho_R$, where the mixing is rather small, $\theta_m~\approx~0$.
As the first approximation, one can neglect the change of
$\theta_m$ in the initial adiabatic stage counting $\theta_m = 0$,
and consider just the vacuum oscillations of \bnue in the
star and on the way from the star to the Earth. In this
case, the
permutation factor coincides with the ``vacuum" permutation factor:
$$
p_{na} \approx p_{vac} \equiv {1\over 2}\sin^2 2\theta~.
\eqno(11)
$$
Consider the effect of the adiabatic transformation
of the neutrinos in the initial stage.
Let $\rho_a$ be the density at which $\kappa = 1$ (see Eq.(5)).
Then, the neutrino flavor changes adiabatically at
$\rho \gg \rho_a$; in the region $\rho \sim \rho_a$, flavor
changes nonadiabatically, and at $\rho \ll \rho_a$, where
$\kappa \gg 1$, one can consider just vacuum oscillations.
Even if the  adiabaticity is restored at
$\rho \ll \rho_R$, the matter effect in this region (especially
at big mixings) is negligibly small. To estimate the effect of
adiabatic and nonadiabatic conversion, one can (simplifying
the picture) consider the
propagation before $\rho_a$ ($\rho \geq \rho_a$) as pure adiabatic
and after $\rho_a$ ($\rho \leq \rho_a$) as strongly nonadiabatic,
i.e. as oscillations in vacuum.
If $\theta_a$ is the mixing angle at $\rho_a$:
 $\theta_a = \theta_m (\rho_a)$,  then the
neutrino state which adiabatically arrives at  $\rho_a$
can be written as
$\nu_a \approx \nu_{1m} \equiv \cos\theta_a \cdot \nu_e - \sin
\theta_a \cdot \nu_\mu$.
Considering vacuum oscillations of $\nu_a$
in the region
$\rho < \rho_a$, as well as on the way from the star to the
Earth, one gets
$$
p_{na} = {1\over 2} \left[1 - \cos2(\theta - \theta_a) \cdot
\cos2\theta\right].
\eqno(12)
$$
The Earth matter effect in the nonadiabatic domain
$(\Delta m^2 < 10^{-9}~ {\rm eV}^2)$ can be neglected due to strong
suppression of mixing.
In the limits of very strong adiabaticity violation
($\rho_a \gg \rho_R$ and
$\theta_a~\cong 0$), Eq.(12) reduces to the
 pure vacuum oscillation result (11). In the opposite case, when
the adiabaticity condition is satisfied everywhere up to
zero densities ($\theta_a~=~\theta$),  Eq.(12) reproduces
the pure adiabatic permutation factor.
At $\rho_a \gg \rho_R$, the condition for $\theta_a$
 can be written as (see Eq.(5)):
$$
\sin ^2 2\theta_a \approx 4\pi \sin 2\theta \cdot {l_{\rho} \over
l_{\nu}}.
\eqno(13)
$$

In the region of the ``just-so" solution,
$\kappa_R \approx 5$, i.e. the adiabaticity condition is strongly violated.
Using Eqs.(12, 13), one finds that at $l_{\rho} = (1 - 3) R_{\odot}$
and $E > 20$ MeV
the permutation factor decreases by $(3 - 5) \%$ in comparison
with the vacuum value. The dependence of the correction on energy
is  very weak.

According to Eqs. (9 - 12), the
adiabatic permutation factor  is always
smaller than the nonadiabatic and
the vacuum  (or strongly nonadiabatic) permutation factors:
$p_a~\leq~ p_{na}~\leq~$  $p_{vac}$. Note that in the
nonresonant case the nonadiabatic transition results in a
stronger effect than the adiabatic transition.
The adiabatic permutation factor can be used to obtain
the lower limit of the permutation effect.

Using the relations in Eqs.(9, 11) we find the upper limits on
$\sin^2 2\theta $ corresponding to
different upper bounds on $p$. In the extreme
cases:
$$
\sin^2 2\theta \leq \left\{\matrix{4 p \cdot (1 - p) \hfill & {\rm (adiabatic,
                       ~ without~ Earth~ effect)} \hfill\cr
2p \hfill & {\rm (strongly~ nonadiabatic)} \hfill\cr}\right.
\eqno(14)
$$
For the upper bounds given in equation (2),
we get the following upper limits on mixing angle at $99\%$ CL:
$$
\sin^2 2\theta \leq \left\{\matrix{0.9 \hfill & (\Delta m^2 \gg
10^{-9} {\rm eV}^2) \hfill\cr
        0.7 \hfill & (\Delta m^2 \ll 10^{-9} {\rm eV}^2) \hfill\cr}\right.
\eqno(15)
$$
These results are exhibited in Figure 3.

In the case of three neutrino mixing, the permutation factor
is determined by the elements of the mixing matrix $U_{ei}$
(i = 1, 2, 3): $p_a = 1 - |U_{e1}|^2$ in the adiabatic limit, and
$p_{na} = 1 - \sum_{i=1,2,3} |U_{ei}|^4$ in the
strongly nonadiabatic limit.

For the resonant channel (neutrino transitions $\nu_{e} - \nu_{\mu}$,
or antineutrino transitions in the case of inverse mass hierarchy),
the permutation factor can be found from the result obtained
above:
$p^{res}_a  = 1 - p_a$. Now $p^{res}_a = \cos^2 \theta$ without
Earth matter effects and the Earth decreases again the transition,
because  now $\theta_m > \theta$ (see Eq.(9)).
In vacuum, $p_{vac}^{res} = p_{vac}$, if there is no CP-violation.
For the resonant
channel, the relation between different permutation factors reads:
$p_a > p_{na} > p_{vac}$.

\vskip 0.8cm

\noindent
{ \bf V. Discussion}

1. We have obtained an upper bound on the permutation
parameter, p, (see Eq.(2))  using  observational  data on the neutrino
burst from SN1987A and  the original neutrino
spectra predicted by neutrino burst models that describe well the
observed luminosity and the burst duration.
We have derived the relation between the permutation factor and the
vacuum mixing angle and have shown that
this relation is practically independent of the structure of the
star in physically interesting regions of neutrino parameter space.
The relation allows one to set upper bounds on the lepton mixing
angle (see Fig.3). The excluded region of neutrino parameters
covers the region of the ``just-so" solution of the solar
neutrino problem, part of the region of the large-mixing-angle
MSW solution,  and part of the region of $\nu_e - \nu_{\mu}$
oscillations which could be responsible for atmospheric
muon neutrino deficit.

2. The upper limit on $p$ derived here can be directly applied to
any transformations of $\bar{\nu}_e$ to  $\bar{\nu}_{\mu}$, or
$\nu_{\mu}$, or $\bar{\nu}_{\tau}$, or
$\nu_{\tau}$ which are independent of, or only weakly depend on, the
neutrino energy.  Spin-flavor conversion,
$\bar{\nu}_{eR} \leftrightarrow \nu_{\mu L}$ (or $\nu_{\tau}$), can result
in spectra permutation with $p$ up to 1/2. This maximal value could
be realized if there is some region inside the star in which the
the interaction with magnetic field, $B$, dominates over the vacuum
and the matter effects: $\mu \cdot B \gg G_{F} \rho /m_{N},
\Delta m^2 /E$, and the neutrino propagates up to this region
adiabatically. The limit on $p$ set in this paper
can be converted to a
limit on the product $\mu \cdot B(r)$, although this restriction
depends sensitively  on the structure of the star.

If neutrino mixing is induced by some flavor off-diagonal
interaction with the ambient medium (``massless oscillations")
[36], then
both neutrino and antineutrino channels can  be resonant.
In this case $p$ may be bigger than 1/2 [37]. The upper bound
set here on $p$
translates into  the upper bounds on the coupling constants of
the new interaction [37].

3. The upper bounds on $p$, and therefore on lepton
mixing, depend strongly on the parameters of the original
neutrino spectra (Fig.2). The integral
characteristics of the neutrino burst (such as total energies
emitted in neutrinos, or the average energy of time
integrated spectra) are determined in large part
by the initial mass of iron core, $M_{Fe}$, and are
independent of most details of the model or of the
explosion mechanism [25]. Since  $M_{Fe}$ and the duration of the
neutrino burst
are  fixed by observations, the integral
parameters of the neutrino burst can, in principle, be
strongly constrained. The difference in fluxes and the average energies of
neutrinos of different species are determined by the known difference in
interactions of these neutrinos. Moreover, the effective temperatures of
neutrino spectra enter as $T^4$ in the luminosity and as $T^5$ in the
interaction rates. This means that small changes in T imply
appreciable changes of other characteristics of the supernova;
this circumstance is reflected in the relatively small spread of
calculated model
parameters (see Table I).

It is of great importance that supernova modelers
 refine their predictions for integral
characteristics of neutrino energy spectra.
One needs to
find the reliable regions, as well as the allowed limits, for parameters
characterizing the energy spectra by making
use of all available information on SN1987A
(excluding, of course, the information on the neutrino burst).

4. Future observations of SN1987A (light curve,
possible manifestations of remnant), when combined with
improvements in the theory of  neutrino transport and  supernova
explosion may
allow one to make stronger inferences.
 The detection
of a new (high-statistics) neutrino burst would make it
possible  to
look for the deviations from simple Fermi-Dirac spectra
modified by a ``chemical potential," especially
the appearance of a high-energy
tail. Confronting  the model calculations
with
data on both the charged current and on the neutral current interactions
(as is possible with SNO, LVD, Superkamiokande) would sharpen the conclusions.

5. We have described  in this paper a method of constraining
lepton  mixing using data on a neutrino
burst produced by gravitational collapse. Depending on
the skepticism of the reader, the results obtained from SN1987A
can be considered as either a demonstration of the method or as
indicating  that  large-angle
lepton mixing is excluded.

\vskip 0.8cm

\noindent
{ \bf Acknowledgements}

We would like to thank E.~Kh.~Akhmedov,
A.~Burrows, A.~Dar, M.~Fukugita,
S.~P. Mikheyev, D.~N.~Schramm, and G.~Senjanovic
 for valuable discussions. One of us (A.S.) is
grateful to H.~Suzuki and H.-T.~Janka for sending their data
and for extensive comments.
Work was supported (A.S.) in part by the National Science Foundation
(grant {\bf NSF $\#$PHY92-45317}) and the Ambrose Monell Foundation.

\vskip 0.8cm
\vfill\eject
\noindent
{ \bf References}

\Rf
[1] R.~Davis, Talk given at the International
Symposium on Neutrino Astrophysics, \break
 Takayama/Kamioka (October 1992)
 K.~S.~Hirata et al., Phys.~Rev.~Lett. {\bf 66}, 9 (1991);
Phys.~Rev. D {\bf 44}, 2241 (1991);
 SAGE: A.~I.~Abazov et al. Phys.~Rev.~Lett. {\bf 67},
 3332 (1991);
 GALLEX: P.~Anselmann et al., Phys.~Lett. B  {\bf 285},
376 (1992).
\Rf
[2]  J.~N.~Bahcall and M.~H.~Pinsonneault,
Rev.~Mod.~Phys. {\bf 64}, 885;  \hbox{I.-J.~Sackmann} et al.,
Astroph.~J. {\bf 360}, 727 (1990);
R. Sienkiewicz J. N. Bahcall, and B. Paczynski, Astroph. J. {\bf 349},
641 (1989);
J.~N.~Bahcall and R.~K.~Ulrich, Rev.~Mod.~Phys. {\bf 60}, 297 (1988);
 S.~Turck-Chieze
 et al., Astroph.~J. {\bf 335}, 415 (1988),
Talk given at the Int.~Conf. ``Neutrino-92";
(1992).
\Rf
[3] V.~N.~Gribov and B.~M.~Pontecorvo, Phys.~Lett.  {\bf 28}, 493 (1967);
J.~N.~Bahcall and
 S.~C.~Frautschi, Phys.~Lett.~B {\bf 29}, 623 (1969);
V.~Barger, R.~J.~N.~Phillips, and K.~Whisnant,
 Phys.~Rev.~D {\bf 24}, 538
(1981);
S.~L.~Glashow and L.~M.~Krauss, Phys.~Lett.~B {\bf 190}, 199
 (1987).
\Rf
[4] V.~Barger, R.~J.~N.~Phillips and K.~Whisnant, Phys.~Rev.~D {\bf 43},
1110 (1991);
A.~Acker,
 S.~Pakvasa and J.~Pantaleone, Phys.~Rev.~D {\bf 43}, 1754
(1991); P.~I.~Krastev and
\hbox{S.~T.~Petcov,} Phys.~Lett.~B {\bf 285}, 85 (1992).
\Rf
[5] S.~P.~Mikheyev and A.~Yu.~Smirnov,  Sov.~J.~Nucl.~Phys. {\bf 42}, 913
(1986); Prog.~Part.~Nucl.
 Phys. {\bf 23}, 41 (1989);
 L.~Wolfenstein, Phys.~Rev.~
D {\bf 17}, 2369 (1978), ibidem, D {\bf 20}, 2634 (1979);
 T.~K.~Kuo and J.~Pantaleone,
Rev.~Mod.~Phys. {\bf 61}, 937 (1989).
\Rf
[6] P.~Anselmann et al., Phys.~Lett. B {\bf 285}, 389 (1992);
 S.~A.~Bludman, D.~C.~Kennedy and P.~G.~Langacker, Phys.~Rev.~D
{\bf 45},
1810 (1992); S.~A.~Bludman, N.~Hata, D.~C.~Kennedy and
P.~G.~Langacker, Pennsylvania preprint UPR-0516T;
 X.~Shi and D.~N.~Schramm, Phys.~Lett.~B {\bf 283}, 305 (1992);
X.~Shi,  D.~N.~Schramm, and
 J.~N.~Bahcall, Phys.~Rev.~Lett.
{\bf 69}, 717 (1992);
 J.~M.~Gelb, W.~Kwong and S.~P.~Rosen, Phys.~Rev.~Lett.~
{\bf 69}, 1846 (1992);
 W.~Kwong and S.~P.~Rosen, Phys.~Rev.~Lett. {\bf 68},
748 (1992);
 A.~Yu.~Smirnov, ICTP preprint IC/92/429;
 P.~I.~Krastev and S.~T.~Petcov, Phys.~Lett. B {\bf 299},
99 (1993);
 L.~M.~Krauss, E.~Gates and M.~White, Phys.~Lett. B {\bf 298},
94 (1993),
 Phys.~Phys.~D {\bf 46}, 1263 (1992),
Phys.~Rev.~Lett. {\bf 70}, 375 (1993).
\Rf
[7] K.~S.~Hirata et al., Phys.~Lett.~B {\bf 280},146 (1992);
 R.~Becker-Szendy et al., Phys.~Rev.~D {\bf 46},
3720 (1992).
\Rf
[8] A.~Zee, Phys.~Lett.~B {\bf 93}, 389 (1980).~
 K.~S.~Baby and E.~Ma, Phys.~Lett.~B {\bf 228},
508 (1989).
 W.~Grimus and H.~Neufeld, Nucl.~Phys. B {\bf 325},
18 (1989).
\Rf
[9] S.~M.~Bilenky and S.~T.~Petcov, Rev.~Mod.~Phys. {\bf 59},671
(1987);
 J.~W.~F.~Valle, Prog.~Part.~Nucl Phys.~
{\bf 26}, 91 (1991).
\Rf
[10] C.~H.~Albright, Phys.~Rev. D {\bf 43}, R3595 (1991);
Phys.~Rev. D {\bf 45}, R725 (1992).
\Rf
[11] A.~Yu.~Smirnov, Proc of the Joint Int. LP-HEP Symposium
(Geneva, 1991),
 S.~Hegarty, K.~Potter and E.~Quercigh, v.~1, 623 (1992).
\Rf
[12] R.~Barbieri, J.~Ellis and M.~K.~Gaillard, Phys.~Lett. B {\bf 90},
249 (1980).
\Rf
[13] E.~Kh.~Akhmedov, Z.~G.~Berezhiani, and G.~Senjanovi$\check c$,
Phys.~Rev.~Lett. {\bf 69}, 3013 (1992).
\Rf
[14] K.~Hirata et al., Phys.~Rev.~Lett. {\bf 58}, 1490 (1987).
\Rf
[15] R.~M.~Bionta et al., Phys.~Rev.~Lett. {\bf 58}, 1494 (1987).
\Rf
[16] E.~N.~Alexeyev et al., Phys.~Lett.~B {\bf 205}, 209 (1988).
\Rf
[17] L.~Wolfenstein, Phys.~Rev. D {\bf 20}, 2634 (1979).
 P.~Reinartz, L.~Stodolsky, Z.~Phys. C {\bf 27}, 507
(1985).
\Rf
[18] S.~P.~Mikheyev and A.~Yu.~Smirnov, Sov.~Phys.~JETP {\bf 64}, 4 (1986);
Proc. of the 6th Moriond Workshop on Massive Neutrinos in Astrophysics
and Particle Physics, (Tignes,
 France) eds O.~Fackler and J.~Tran
Than Van, 355 (1986); G.~M.~Fuller et al., Astroph.~J.
{\bf 322}, 795 (1987); G.~M.~Fuller et al, Astroph.~J.
{\bf 389}, 517 (1992).
\Rf
[19] L.~Wolfenstein, Phys.~Lett. B {\bf 194}, 197 (1987).
\Rf
[20] A.~Yu.~Smirnov, talk given at XXth  ICCR, Moscow 1987 (unpublished),
see in [5].
\Rf
[21] J.~A.~Frieman, H.~E.~Haber, and K.~Freese, Phys.~Lett.~B {\bf 200},
115 (1988).
\Rf
[22] A.~E.~Chudakov, Ya.~S.~Elensky and S.~P.~Mikheyev,
Proc. of the NATO Advanced Study
 Institute ``Cosmic Gamma Rays,
Neutrinos and Related Astrophysics" (Erice, Italy,
 1988) Kuwer Acad. Publ., ed by M.~M.~Shapiro and J.~P.~Wefel
p.~131 (1989).
\Rf
[23] V.~S.~Imshennik and D.~K.~Nadjozhin, Sov.~Astrophys.  and
Space Phys.~Rev.
{\bf 8}, 1 (1989).
 ed. R.~A.~Sunyaev.
\Rf
[24] S.~E.~Woosley, J.~R.~Wilson, and R.~Mayle, Astroph.~J.
{\bf 302}, 19 (1986).
\Rf
[25] R.~Mayle J.~R.~Wilson, and D.~Schramm, Astroph.~J.
{\bf 318}, 288 (1987).
 J.~Wilson, Nucl.~Phys.
B (Proc. Suppl.) {\bf 13}, 380 (1990).
\Rf
[26] S.~W.~Bruenn, Phys.~Rev.~Lett. {\bf 59}, 938 (1987).
\Rf
[27] P.~M.~Giovanoni, P.~C.~Ellison, and S.~W.~Bruenn,
Astroph.~J. {\bf 342}, 416 (1989).
\Rf
[28] A.~Burrows, Astroph.~J. {\bf 334}, 891 (1988).
\Rf
[29] E.~S.~Myra and A.~Burrows, Astroph.~J. {\bf 364} 222 (1990).
\Rf
[30] A.~Burrows, D.~Klein, and R.~Gandhi, Phys.~Rev. D {\bf 45},
3361 (1992).
\Rf
[31] H.~-T.~Janka and W.~Hillebrandt, Astron.~Astroph.~Suppl.
{\bf 78}, 375 (1989);
Astron.~Astroph. {\bf 224}, 49 (1989).
\Rf
[32] H.~-T.~Janka Ph.~D.~thesis, MPA-preprint 587,
Max-Plank-Institut fur Astrophysik (1991)
 Proc. of the Int.~Workshop on ``Frontier Objects in
Astrophysics and Particle Physics,"
 Vulcano, Italy, 1992,
to be published by the Societa Italiana di Fisica, eds.
 F.~Giovanneli and G.~Mannocchi; MPA-preprint 720.
\Rf
[33] H.~Suzuki, Proc. of Int. Workshop on Unstable Nuclei in
Astrophysics, ed. S.~Kubono
 and T.~Kajino, (World Scientific, 1992)
p. 281; talk given at the Int. Symp. on
 Neutrino Astrophysics,
Takayama/Kamioka (October 1992).
\Rf
[34] W.~Hillebrandt and R.~G.~Wolff in ``Nucleosynthesis - Challenges and
New developments,"
 W.~D.~Arnett and J.~W.~Truran (eds.) (The
Univ. of Chicago Press, Chicago, 1985) p.~131.
\Rf
[35] K.~Nomoto and M.~Hashimoto, Phys.~Rep. {\bf 163}, 13 (1988).
\Rf
[36] L.~Wolfenstein, see in [5], M.~M.~Guzzo, A.~Masiero and
S.~T.~Petcov, Phys.~Lett.~B {\bf 260},
 154 (1991); V.~Barger, R.~J.~N.~Phillips
and K.~Whisnant, Phys.~Rev. D {\bf 44} 1629 (1991).
\Rf
[37] P.~I.~Krastev, S.~Petcov and A.~Yu.~Smirnov, in preparation
(see Proc. of the 4th Int.~Symp.~
 on ``Neutrino Telescopes,"
Venice, March 1992).

\vfill\eject

\noindent
{\bf Figure Captions}

Figure 1. The cumulative energy spectrum of the events in the
Kamiokande and IMB detectors predicted for different values
of permutation  factor $p$ (from left to right: $p$ = 0, 0.1, 0.2, 0.3,
0.4 and 0.5). Original
spectra were taken  according to the model [24]. Histogram
shows the detected spectrum.

Figure 2. Upper bounds on permutation factor from SN1987A
data as functions of the original spectra parameters.
a). The dependence of the upper bound (99$\%$ CL) as a function
of the average muon energy $\bar{E}_{\mu}$ on the ratio of total
energies, $r$,
emitted  in \bnum and \bnue.
Other parameters  are fixed as follows: $\bar{E}_e = 13$ MeV,
$\eta_e = \eta_{\mu} = 2$. Top to bottom: $r$ = 0.4, 0.6, 0.8, 1.0,
1.2. b). The dependence of the same
bound as in a) on the electron antineutrino
energy,  $\bar{E}_e$.
Other parameters: $r = 1$, $\eta_e = \eta_{\mu} = 2$. Top to bottom:
$\bar E_e$ = 12, 10, 13, 14, 16 MeV.
c). The dependence of the same bound as in a) on the spectrum distortion
parameter $\eta_{\mu}$. Top to bottom: $\eta_\mu$ = 0, 1, 2, 3, 4.
Other parameters:
$\bar{E}_e = 13$ MeV, $r = 1$,
$\eta_e = 2$. d). The same dependence as in c) for $\eta_e = 0$.
Top to bottom: $\eta_\mu$ = 0, 1, 2, 3, 4.  e). The upper bound
on $p$ at different
confidence levels. Top to bottom: $p$ =
99.9\%, 99.5\%, 99\%, 98\%, 95\%.

Figure 3.(a) The excluded regions of the neutrino parameters
for different upper bounds on permutation factor (figures
at the curves). In the region $\Delta m^{2} = (10^{-8} -
10^{-9})$ eV$^2$ the restrictions (shown by dashed lines)
may appreciably depend on density distribution in the
star. In the region $\Delta m^2 = (10^{-5} - 10^{-6})$ eV$^2$
the approximation  $p = constant$ does not work due to
Earth matter effect (dotted lines). Also the regions of the
solar neutrino problem solutions by the MSW effect and
``just-so" oscillation, as well as the region responsible for
atmospheric muon neutrino deficit are shown. Shadowed curve
depicts the upper bound on neutrino parameters from the
reactor oscillation experiments.  (b) This figure shows
the excluded region as a function of $<p>$, where $<p>$
is averaged over the electron distribution function. The contours are
for $<p>$ = 0.05, 0.1, 0.15, 0.2, 0.25, 0.3, 0.35 and 0.4.

Figure 4. The upper bounds on the mixing angle in the region of a strong
Earth effect. The curves are shown for 95$\%$ and 99$\%$ CL. The
original spectra by MWS [25] are used.

\vfil\eject
\noindent
{\bf Table 1.} Integral characteristics of the neutrino bursts
in different models and corresponding upper bounds on permutation
parameters. (MWS - [24], Bruenn - [26], Burrows - [28 - 30],
Janka - [31, 32]).
\vskip 0.8cm
\vbox{\offinterlineskip
\halign{\strut \vrule \hfil \quad #\quad \hfil
              &\vrule \hfil \quad #\quad \hfil
              &\vrule \hfil \quad #\quad \hfil
              &\vrule \hfil \quad #\quad \hfil
              &\vrule \hfil \quad #\quad \hfil
              &\vrule \hfil \quad #\quad \hfil
              &\vrule \hfil \quad #\quad \hfil
              &\vrule \hfil \quad #\quad \hfil \vrule \cr
\noalign{\hrule}
Model & $\bar{E}_{e}, $ & $\bar{E}_{\mu}, $ & r &
$\eta_e$ & $\eta_{\mu}$ & $p$, & $p$, \cr
    & MeV & MeV &     &     &   & 95$\%$ CL & 99$\%$ CL \cr
\noalign{\hrule}
MWS & 13.8  & 22.3  & 0.9 & 3.8  & 0.6 & 0.27 & 0.42 \cr
Bruenn & 13  & 25  & 0.8 &  2  & 3 & 0.18 & 0.27 \cr
Bruenn & 13  & 25  & 0.8 &  2  & 2 & 0.17 & 0.26 \cr
Burrows & 11.1  & 21  & 1.0 & 0.8  & 2 & 0.24 & 0.38  \cr
Janka & 14  & 22  & 0.8 &  2  & 2 & 0.23 & 0.35  \cr
\noalign{\hrule}
}}

\end